\documentclass[12pt]{article}

\usepackage{amssymb}
\usepackage{amsmath}
\usepackage{graphics}
\usepackage{array}
\usepackage{afterpage}
\usepackage{epsfig}
\usepackage{float}

\begin{document}
\title{Transport Theoretical Approach to the Nucleon Spectral Function in
Nuclear Matter\footnote{Work supported by BMBF, DFG and GSI Darmstadt}}
\author{J. Lehr, M. Effenberger, H. Lenske, S. Leupold and U. Mosel\\
Institut f\"ur Theoretische Physik,
Universit\"at Giessen,\\
Heinrich-Buff-Ring 16, D-35392 Giessen, Germany\\
UGI-00-3}
\maketitle
\date{}

\begin{abstract}
The nucleon spectral function in infinite nuclear matter is
calculated in a quantum transport theoretical approach. Exploiting
the known relation between collision rates and correlation functions
the spectral function is derived self-consistently. By re-inserting
the spectral functions into the collision integrals the description
of hard processes from the high-momentum components of wave
functions and interactions is improved iteratively until
convergence is achieved. The momentum and energy distributions and
the nuclear matter occupation probabilities are in very good
agreement with the results obtained from many-body theory.
\end{abstract}

\noindent PACS numbers: 21.65.+f, 24.10.Cn

\noindent {\it Keywords}: nuclear matter, many-body theory, nucleon
spectral function
\bigskip

\noindent A longstanding problem of nuclear many-body theory is the
question to what extent short-range correlations are contributing to the
properties of nuclear matter. While the bulk properties of nuclear
matter are mainly affected by long-range mean-field dynamics, the picture,
however, changes if nuclei are probed at large energy and momentum
transfer. The spectral functions, obtained for example from $A(e,e'p)X$
experiments \cite{Witt90}, are in energy and momentum much wider spread
than predicted by mean-field dynamics. Obviously, the high momentum and
energy structure of spectral functions has important consequences for
dynamical processes, e.g. sub-threshold particle production on nuclei.

An overall measure of short-range correlations in nuclear matter is
the depletion of occupation probabilities in ground state momentum
distributions by about 10\%. The processes behind this number are
such that states from inside the Fermi sphere are scattered into
high momentum configurations which clearly are not of mean-field
nature. As a result, a momentum distribution with a long high
momentum tail is generated extending much beyond the Fermi surface.
An important finding is that the magnitude and the shape of the
high momentum component is almost independent of the system under
consideration while the inner parts, especially in light nuclei,
are affected by the shell structure and finite size effects. Hence,
the high momentum tails of the spectral functions are likely to
reflect a universal property of nuclear many-body dynamics at
short distances.

Theoretically, many attempts have been made to understand short-range
correlations in nuclei. Approaches based on nuclear many-body theory up
to explanations referring to the QCD aspects of strong interactions
\cite{Strikman} have been proposed. Obviously, before conclusions on
non-standard phenomena can be drawn the many-body theoretical aspects of
short-range correlations must be understood in detail. In fact, the
results obtained from many-body theory are describing the available data
rather satisfactorily. In recent years the theoretical results have
converged, at least for the depletion of nuclear matter occupation
probabilities. The majority of the model calculations are using
Brueckner and Dirac-Brueckner techniques, see e.g.\
\cite{Ramos,FP84,MKP95,DL96,DL97}. In \cite{Peter} a correlation
dynamical treatment was applied. Most of the approaches use the
quasi-particle approximation, i.e.\ a sharp energy distribution for the
nucleons is assumed (see e.g.\ \cite{FdJM}). Occupation probabilities
in finite nuclei could be well described in a second RPA approach
\cite{EL89} and by polarization self-energies \cite{LW90}, respectively.
The Dirac-Brueckner calculations in \cite{DL97}, including hole-hole
propagation, led to an extended and numerically rather involved
energy-momentum structure of self-energies. However, the net effect on
binding energies and occupation probabilities was only moderate, but
improving the agreement with empirical data.

Here, we investigate spectral functions in nuclear matter by quantum
transport theory \cite{KB62,BM90}, thereby taking up a proposal of
Danielewicz and Bertsch \cite{DB}. Indeed, the present study is
motivated by our recent implementation of off-shell effects in a
transport theoretical treatment of heavy-ion and other nuclear
collisions \cite{ef_off,ef_dilep}.  A sound theoretical basis how to 
treat off-shell effects in transport equations
has been given in \cite{leupold} (see also
\cite{cj}). A central result of transport theory is that collision rates
and correlation functions are directly related: The calculation of
either of the two quantities requires the knowledge of the other one.
Theoretically, this corresponds to a rather involved self-consistency
problem for which a direct solution apparently does not exist. Similar
to \cite{DL97}, a practical approach is obtained by an iterative
procedure. Successive approximations for self-energies, spectral
functions and collision integrals are obtained be re-inserting the
corresponding quantities from previous cycles of the calculation until
convergence is achieved. The method is discussed below. Results
are presented and compared to the work of Benhar et
al. \cite{benhar} who calculated the nuclear matter spectral function
within the framework of correlation-basis theory and of Ciofi degli Atti
et al. \cite{C90,C95} who derived a global parameterization of spectral
functions for finite nuclei and nuclear matter. 
\medskip

In quantum transport theory \cite{KB62,BM90} dynamical
processes are described by the one-particle correlation functions
\begin{eqnarray}\label{corrf}
g^>(1,1')&=&-i\langle \Psi(1)\Psi^\dagger(1')\rangle \nonumber \\
g^<(1,1')&=&i\langle \Psi^\dagger(1')\Psi(1)\rangle,
\end{eqnarray}
where $\Psi$ are the nucleon field operators in Heisenberg
representation. They account for the non-stationary processes which
introduce a coupling between causal and anti-causal single particle
propagation. In other words, states from below and above the
Fermi surface are dynamically mixed as discussed above.
Correspondingly, in an interacting quantum system the
single particle self-energy operator includes correlation
self-energies $\Sigma^{<>}$ which couple particle and hole degrees
of freedom \cite{KB62,BM90}. Obviously, $g^{<>}$ and $\Sigma^{<>}$
are closely related to each other and must be determined by those
parts of the fundamental interactions producing non-stationary
effects. The wanted relation is obtained from transport theory.
After a Fourier transformation to energy-momentum representation
one finds for the self energies \cite{KB62}
\begin{eqnarray}
  \label{eq:sigma>}
\Sigma^>(\omega,p)&=&g\int{d^3p_2d\omega_2\over(2\pi)^4}
{d^3p_3d\omega_3\over
(2\pi)^4}{d^3p_4d\omega_4\over(2\pi)^4}(2\pi)^4\delta^4(p+p_2-p_3-p_4)
\, \overline{\vert{\cal{M}}\vert^2}   \nonumber \\
    & &{} \times g^<(\omega_2,p_2)g^>(\omega_3,p_3)g^>(\omega_4,p_4)
\end{eqnarray}
\begin{eqnarray}
  \label{eq:sigma<}
\Sigma^<(\omega,p)&=&g\int{d^3p_2d\omega_2\over(2\pi)^4}
{d^3p_3d\omega_3\over
(2\pi)^4}{d^3p_4d\omega_4\over(2\pi)^4}(2\pi)^4\delta^4(p+p_2-p_3-p_4)
\, \overline{\vert{\cal{M}}\vert^2} \nonumber \\
    & &{} \times g^>(\omega_2,p_2)g^<(\omega_3,p_3)g^<(\omega_4,p_4),
\end{eqnarray}
where $g=4$ is the spin-isospin degeneracy factor and $\overline{\vert{\cal{M}}
\vert^2}$ denotes the square of the nucleon-nucleon scattering amplitude,
averaged over spin and isospin of the incoming nucleons and summed over spin
and isospin of the outgoing nucleons. Since
both $g^{<>}$ and $\Sigma^{<>}$ describe the correlation dynamics,
the spectral function can be obtained from either of the two
quantities as the difference over the cut along the energy real
axis. In terms of the correlation propagators,
\begin{equation}\label{spf}
a(\omega,p)=i\left( g^>(\omega,p)-g^<(\omega,p) \right).
\end{equation}

In a non-relativistic formulation the single particle spectral
function is found explicitly as
\begin{equation}
 \label{eq:spectral}
a(\omega,p)={\Gamma(\omega,p)\over(\omega-{p^2\over 2 m_N}-
\textrm{Re}\Sigma(\omega,p))^2+\frac{1}{4}\Gamma^2(\omega,p)},
\end{equation}
including the particle and hole nucleon self-energy $\Sigma$, which
accounts for long-range mean-field dynamics. The width of the
spectral distribution is determined by the imaginary part of the
self-energy,
\begin{equation}
  \label{eq:gamma}
  \Gamma(\omega,p)=2\textrm{Im}\Sigma(\omega,p)=i(\Sigma^>(\omega,p)-
\Sigma^<(\omega,p)).
\end{equation}
From Eqs. (\ref{eq:sigma>}), (\ref{eq:sigma<}) it is apparent that the high
momentum, i.e.\ short range, components of nuclear interactions are of primary
importance for the energy-momentum spreading of the single particle
strength. Obviously, the $\delta$-shaped quasi-particle
distribution is recovered for $\textrm{Im}\Sigma \to 0$, i.e.\
vanishing correlations.

The correlation propagators $g^{<>}$ are given by
\begin{align}
  \label{eq:g<>}
    g^<(\omega,p) &= i a(\omega,p)f(\omega,p), \\
    g^>(\omega,p) &= -i a(\omega,p)(1-f(\omega,p))
\end{align}
in terms of the energy-momentum phase space distribution function
$f(\omega,p)$. Since we are dealing with a system at $T=0$, $f$ reduces to
\begin{equation}
f(\omega,p)=\Theta(\omega_F-\omega)
\end{equation}
with the Fermi energy $\omega_F$. Therefore,
\begin{align}
&\Sigma^>(\omega,p)=0,\quad \Gamma(\omega,p)=-i\Sigma^<(\omega,p)\quad
\textrm{for}\quad \omega\le \omega_F \\
&\Sigma^<(\omega,p)=0,\quad \Gamma(\omega,p)=i\Sigma^>(\omega,p)\quad
\textrm{for}\quad \omega\ge \omega_F.
\end{align}

In order to obtain $\Gamma$, we have to calculate the self-energies
$\Sigma^>$ and $\Sigma^<$. Since these quantities themselves depend
on $\Gamma$ via the spectral function $a(\omega,p)$, the
calculation has to be done iteratively.
\medskip

Diagrammatically, the particle- and hole-type transition rate $\Sigma^>$
and $\Sigma^<$, Eqs. (\ref{eq:sigma>}) and (\ref{eq:sigma<}), are of
two-particle--one-hole (2p1h) and one-particle--two-hole (1p2h)
structure, respectively. In this respect, they are of the same basic
structure as the polarization self-energies considered  in many-body
theoretical descriptions. However, while in many-body theory the
polarization self-energies are typically included perturbatively in
lowest order only by performing the integrations over intermediate 2p1h
and 1p2h states with quasi-particle spectral functions, e.g. in Ref.
\cite{benhar} and also Refs. \cite{EL89,LW90}, we apply a more extended
scheme accounting for higher order effects.

A first attempt to extend the many-body scheme to higher order
calculations was made in \cite{DL97}. Since, as expected, the
computational effort was found to increase considerably approximations
on spectral functions of particle intermediate states had to be invoked.
An important aspect of the present transport theoretical approach is the
fully self-consistent treatment of particle and hole strength function
in each stage of the calculation. Within our iterative approach this is
achieved by calculating the correlation self-energies with the
self-consistently obtained spectral functions. Hence, the converged
results include a non-perturbative summation of a whole series of
$n$p $m$h diagrams.

Dynamically, the strength of the transition rates is determined by
the in-medium (off-shell) nucleon-nucleon scattering amplitude
$\cal{M}$. We now make the extreme assumption of a constant
transition amplitude, as is appropriate for the short-range part of
the $NN$ interaction. Neglecting the energy-momentum dependence of
the amplitude clearly simplifies the computations. Then, $\cal{M}$
contributes only as an overall multiplicative factor to the
transition rates which, at a given density, may be treated as an
empirical parameter $\overline{\cal{M}}$. The self-energies
$\Sigma^>$ and $\Sigma^<$ are then given by
\begin{eqnarray}
  \label{eq:sigma>_2}
\Sigma^>(\omega,p)&=&-4i{\overline{\vert{\cal{M}}\vert^2}\over(2\pi)^6}
\int d\omega_3\int d\omega_2\int dp_3p_3^ 2\int dp_2p_2^2 
{d\cos\vartheta_2\over p_{\textrm{tot}}p_3}a(\omega_2,p_2)
\nonumber\\ & & 
\hspace{-1cm}\times f(\omega_2,p_2)a(\omega_3,p_3)(1-f(\omega_3,p_3)) \int
dp_4 p_4 a(\omega_4,p_4)(1-f(\omega_4,p_4))\nonumber \\ & &
\end{eqnarray}
and
\begin{eqnarray}
  \label{eq:sigma<_2}
\Sigma^<(\omega,p)&=&4i{\overline{\vert{\cal{M}}\vert^2}\over(2\pi)^6}
\int d\omega_3\int d\omega_2\int dp_3p_3^2\int dp_2p_2^2 \int
{d\cos\vartheta_2\over p_{\textrm{tot}}p_3} a(\omega_2,p_2)\nonumber\\
& & {} \times \! (1-f(\omega_2,p_2))a(\omega_3,p_3)f(\omega_3,p_3) 
\int
dp_4 p_4a(\omega_4,p_4)f(\omega_4,p_4)\nonumber \\ & &
\end{eqnarray}
with $p_{\textrm{tot}}=\vert\vec p+\vec p_2\vert$, where four
integrations due to the delta function and two over the azimuthal
angles $\varphi_2,\varphi_3$ already have been carried out. The
remaining six dimensional integrals were calculated on a
$(\omega,p)$-grid with boundaries $\vert\omega\vert\le0.5$ GeV and
$p\le p_{\textrm{max}}=1.25$ GeV and a mesh size of 5 MeV in both
directions. 

In the present study we work in infinite nuclear matter at equilibrium
with the Fermi momentum $p_F = 1.33$ fm$^{-1}$, density
$\rho$=0.16 $\textrm{fm}^{-3}$ and a binding energy per particle of
$\omega_F=-16$~MeV. The real part of the self-energy Re$\Sigma$ 
is chosen to be independent of momentum and energy. In fact, a constant
Re$\Sigma$ only serves to define the scale for the excitation energy
$\omega$ which in our case is given by $\omega\geq \omega_F$. The 
energy and momentum dependent pieces of Re$\Sigma$ play a more important
role because they will modify the pole structure of the propagators.
Inside the Fermi sphere typically a quadratic momentum dependence is
found leading to a scaling of the kinetic energy by an effective mass
and a global compression of spectra will be found. Hence, leaving out
the momentum dependence will affect the final results only to a minor
degree. Neglecting the energy dependence might introduce larger
uncertainties because this amounts to a violation of analyticity as expressed
by the dispersion relation between real and imaginary parts of
polarization self-energies. Since we do obtain an energy dependent
imaginary part analyticity could in principle be restored by calculating
Re$\Sigma$ dispersively. However, using a constant, energy independent
transition matrix element clearly fails for large positive energies
because the imaginary part would continue to increase beyond any limit
with the level density of two-particle states. Thus, the imaginary part
of the self-energy is not available over the full range of energies
necessary for solving the dispersion equation. Rather than introducing
an energy cut-off as an additional free model parameter we decide at the
present stage to completely neglect the energy dependence of Re$\Sigma$.

\begin{figure}[H]
  \begin{center}

\epsfig{file=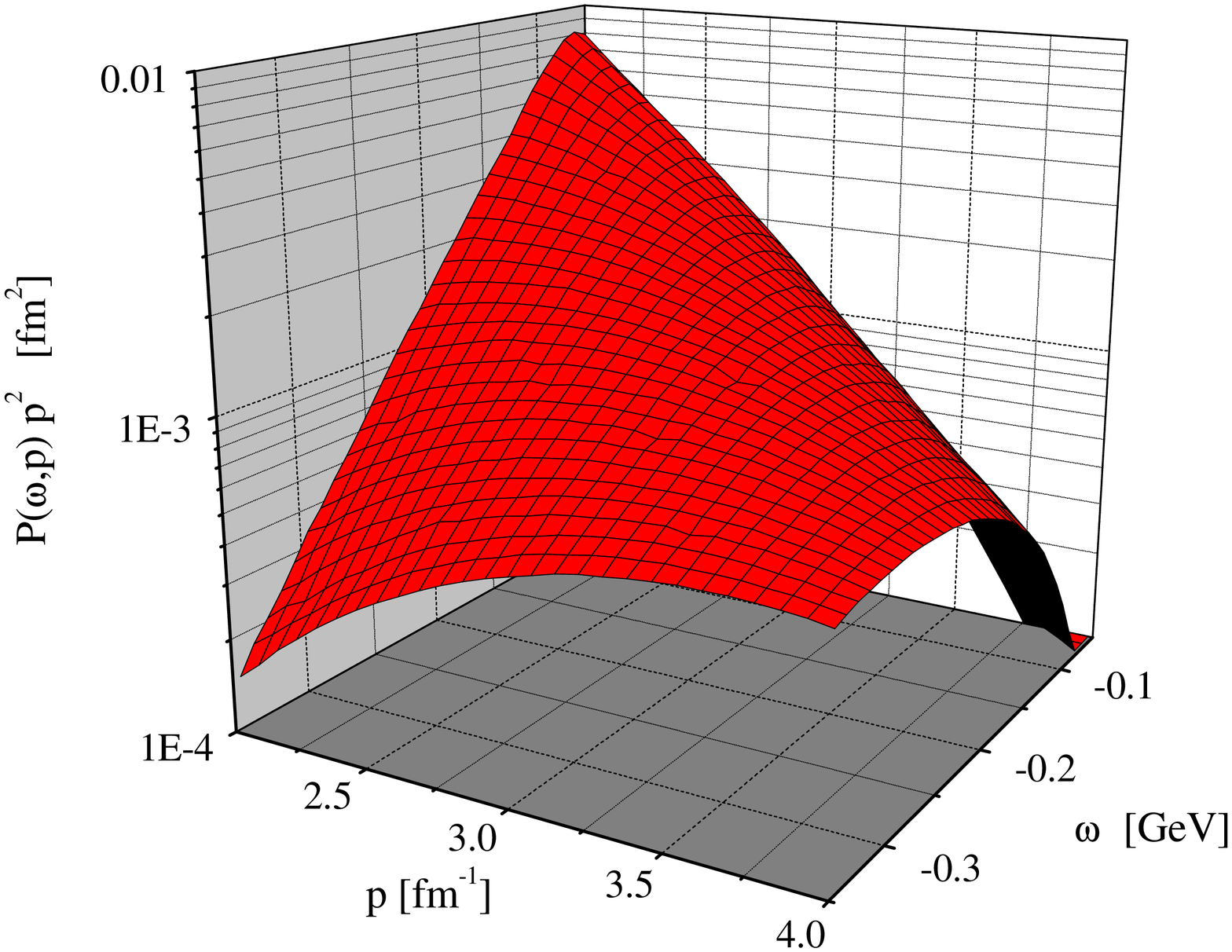,width=14cm}
    \caption{Energy and momentum dependence of the nucleon spectral function
for energies below $\omega_F$.}\label{fig1}
  \end{center}
\end{figure}

The average amplitude was adjusted such that the spectral functions
obtained with many-body theoretical methods by Benhar et
al. \cite{benhar} are reproduced leading to
$\left(\overline{\vert{\cal{M}}\vert^2}\right)^{1/2}$=207 MeV fm$^{3}$.
Relating this value to on-shell processes would correspond to a constant
cross section of about 20~mb. 
In Fig.\ \ref{fig1} the full
energy-momentum structure of the resulting nucleon spectral function
\begin{equation}
  \label{eq:spec_p}
  P(\omega,p) = N\ a(\omega,p)
\end{equation}
is displayed, where $N$ is a normalization constant.
The same normalization as in the work
of Benhar et al. \cite{benhar} is used.
Our results are in remarkable agreement with the calculations of Ciofi degli 
Atti et al. \cite{C95} as seen by comparing to Fig.~10 in that reference.

\begin{figure}[H]
  \begin{center}
\epsfig{file=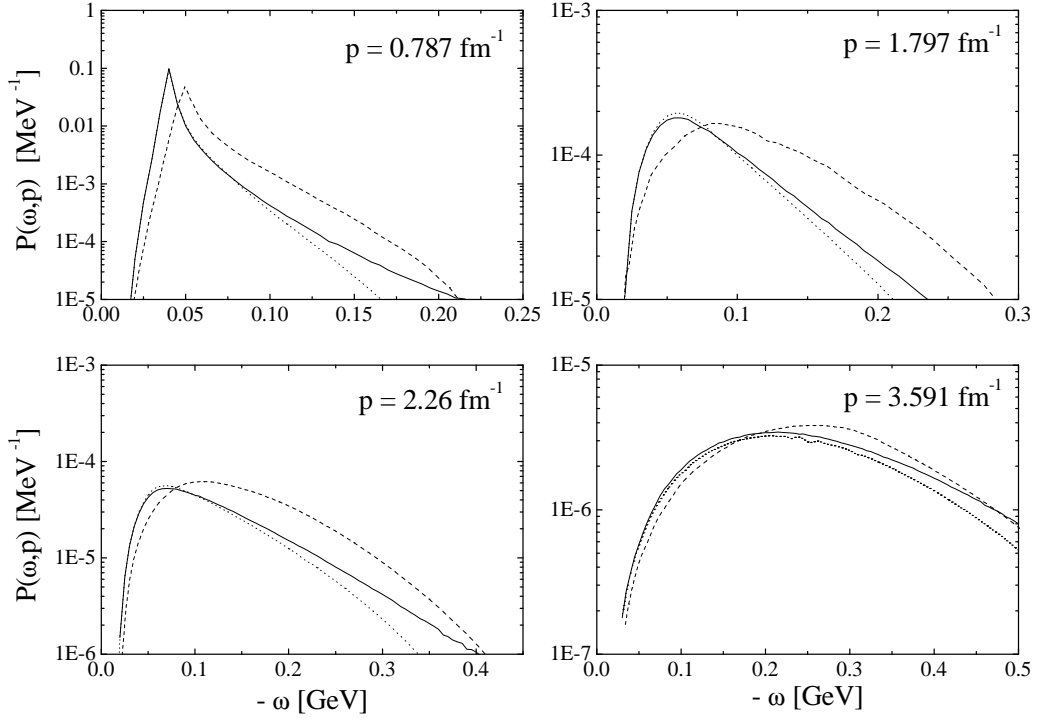,width=14cm}
    \caption{Nucleon spectral function for different 
constant momenta for energies below $\omega_F$.
The dotted curves have been obtained after the first iteration step using
a constant initial width of 0.1 MeV (see text).
The dashed curves show the results of Benhar et al. \cite{benhar,fantoni}.}
\label{mom_cuts}
  \end{center}
\end{figure}

For a more quantitative comparison cuts for several momenta are shown
in Fig.\ \ref{mom_cuts} as a function of energy $\omega$ below the
Fermi surface and compared to the results of Ref. \cite{benhar}. The
agreement is surprisingly good considering the seemingly drastic
approximation invoked for $\cal{M}$ and the neglect of analyticity.

We conclude from this agreement that the spectral function of nuclear
matter in the hole-sector is rather insensitive to the specific energy
and momentum structure of the interaction once a fully self-consistent
calculation is performed.  The effect of self-consistency is illustrated in 
Fig. 2
where spectral functions obtained after the first iteration are compared
to the fully converged results. The iteration scheme was initialized by 
choosing
spectral functions with a constant width of  0.1~MeV as start value. 
On a level of about 30\% global features, as e.g. the location and
height of the peak structures, are described already by  the first iteration.
At first sight, this seems to support the pertubative approach usually applied
in many-body theoretical approaches. However,
during the self-consistency cycle strength is re-distributed from the peak 
into the tail regions. Hence,  higher order polarization effects act in a 
similar way as
increasing the effective coupling strength. The results show that 
the full transport theoretical calculation apparently leads to a 
self-consistent adjustment 
of spectral functions which ultimately is dominated by phase space effects. 

\begin{figure}[H]
  \begin{center}
\epsfig{file=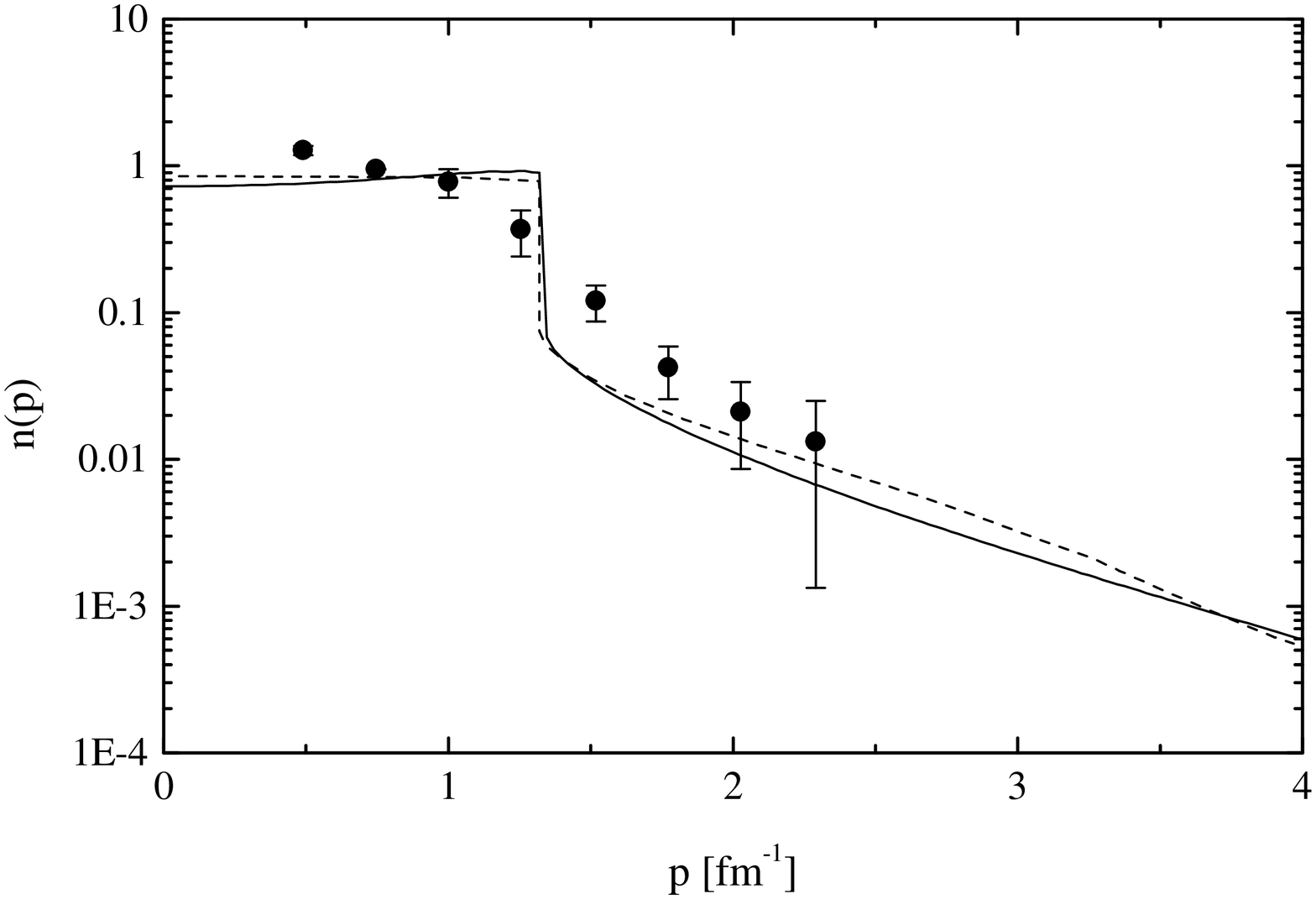,width=14cm}
    \caption{Momentum distribution in nuclear matter calculated from the
spectral function. The dashed curve is the result from Benhar et al.
\cite{benhar}. The data are from \cite{cda_dat}.}\label{mom_distr}
  \end{center}
\end{figure}

The agreement with the many-body calculations of \cite{benhar} extends
also to the single nucleon momentum distribution in nuclear matter
defined as
\begin{equation}
  \label{eq:mom_distr}
  n(p)=\int_{-\infty}^{\omega_F}d\omega P(\omega,p).
\end{equation}
Results are displayed in Fig. \ref{mom_distr}. Again the agreement is close
to perfect except for the increase of the distribution towards
$p_F$, which is in contrast to the calculations of \cite{benhar}.
Calculations in a schematic model using an empirical form for
$\Gamma(\omega,p)$ with parameters given in Ref. \cite{Mahaux} show that
this behaviour is a direct consequence of the missing analyticity in the
self-energies. The schematic approach also confirms that the momentum
dependence of Re$\Sigma$ indeed leads to minor effects in the bulk
behaviour of the spectral functions except close to the Fermi surface.
The momentum distribution is affected mainly in the extreme momentum
tail where a steeper slope is found. In this context it is important
to note that $n(p)$ must decrease stronger than $\cal{O}$$(\frac{1}{p^4})$
in order to
have a convergent result for the kinetic energy distribution of the
correlated ground state. The transport theoretical results in
Fig. \ref{mom_distr} fulfill this constraint.

Finally, we remark that the shape of the momentum distribution is
closely related to the value of $\overline{\cal{M}}$: Increasing the
value - corresponding to a stronger interaction amongst the nucleons -
would increase the occupation of  states above $p_F$ and soften the
Fermi edge, leading to a better description of the data in Fig. 
\ref{mom_distr} for $p>p_F$.

\bigskip

Summarizing, correlations in nuclear matter have been described by an approach
that allows to go beyond the perturbative level to which conventional
many-body methods are constrained in practice. In a first application,
the transition rates were calculated with a momentum independent
nucleon-nucleon scattering amplitude. That a transport theoretical
description with a simple approximation for the scattering amplitude
could reproduce the momentum distribution obtained in state-of-the-art
many-body calculations had already been observed by the authors of
\cite{DB}. The results here show in addition that also the spectral
functions are dominated by phase space effects rather than by the
off-shell momentum structure of interactions. The excellent agreement
of the nucleon spectral function at far-off-shell energies and momenta
with those obtained in sophisticated state-of-the-art nuclear many-body
calculations also provides some a-posteriori justification for the
method to include off-shell effects in nuclear transport calculations
developed in Ref.\ \cite{ef_off}.
\vspace{1cm}

\noindent Acknowledgements:\\
\noindent We are grateful to S. Fantoni for providing us with his
spectral functions and to C. Ciofi degli Atti for helpful comments.

%%%%%%%%%%%%bibliography%%%%%%%%%%%%%%%%%%%%%%

\newpage

%%%%%%%%%%%%%figures%%%%%%%%%%%%%%%%%%%%%%%%%%%%%%%%

\newpage

\newpage

\begin{thebibliography}{99}
\bibitem{Witt90}for a review see, e.g., P.K.A. de Witt Huberts, J. Phys. G
(Nucl. Part. Phys.) {\bf 16} (1990) 507.
\bibitem{Strikman}L. Frankfurt and M.I. Strikman, Phys. Rep. {\bf 76} (1981) 215.
\bibitem{Ramos}A. Ramos, A. Polls and W.H. Dickhoff, Nucl. Phys. {\bf A503} (1990) 1;
A. Ramos, A. Polls and W.H. Dickhoff, Phys. Rev. {\bf C43} (1991) 2239.
\bibitem{FP84}S. Fantoni and V.R. Pandharipande, Nucl. Phys. {\bf A427} (1984) 473.
\bibitem{MKP95}H. M\"uther, G. Knehr and A. Polls, Phys. Rev. {\bf C52} (1995) 2955.
\bibitem{DL96}F. de Jong and H. Lenske, Phys. Rev. {\bf C54} (1996) 1488.
\bibitem{DL97}F. de Jong and H. Lenske, Phys. Rev. {\bf C56} (1997) 154.
\bibitem{Peter}A. Peter, W. Cassing and J.M. Haeuser, Nucl. Phys. {\bf A573} (1994) 93. 
\bibitem{FdJM} F. de Jong and R. Malfliet, Phys. Rev. {\bf C44} (1991)
998.
\bibitem{EL89}F.J. Eckle, H. Lenske, G. Eckle, G. Graw, R. Hertenberger,
H. Kader, F. Merz, H. Nann, P. Schiemenz and H.H. Wolter, 
Phys. Rev. {\bf C39} (1989) 1662;
F.J. Eckle, H. Lenske, G. Eckle, G. Graw, R. Hertenberger, H. Kader, 
H.J. Maier, F. Merz, H. Nann, P. Schiemenz and H.H. Wolter, Nucl. Phys. 
{\bf A506} (1990) 159.
\bibitem{LW90}H. Lenske and J. Wambach, Phys. Lett. {\bf B249} (1990) 377.
\bibitem{KB62}L.P. Kadanoff and G. Baym, Quantum Statistical Mechanics
(Benjamin, New York, 1962).
\bibitem{BM90} W. Botermans and R. Malfliet, Phys. Rep. {\bf 198} (1990) 115.
\bibitem{DB}G. Bertsch and P. Danielewicz, Phys. Lett. {\bf B367} (1996) 55.
\bibitem{ef_off}M. Effenberger and U. Mosel, Phys. Rev. {\bf C60} (1999) 
051901.
\bibitem{ef_dilep}M. Effenberger, E.L. Bratkovskaya and U. Mosel, Phys. Rev.
{\bf C60} (1999) 044614.
\bibitem{leupold} S. Leupold, nucl-th/9909080; Nucl. Phys. {\bf A}, in print.
\bibitem{cj} W. Cassing and S. Juchem, Nucl. Phys. {\bf A665} (2000) 385; 
nucl-th/9910052, Nucl. Phys. {\bf A}, in print.
\bibitem{benhar} O. Benhar, A. Fabrocini and S. Fantoni, Nucl. Phys.
{\bf A505} (1989) 267; O. Benhar, A. Fabrocini and S. Fantoni, Nucl. Phys. 
{\bf A550} (1992) 201.
\bibitem{C90}C. Ciofi degli Atti, S. Simula, L.L. Frankfurt and M.I. 
Strickmann, Phys. Rev. {\bf C44} R7 (1991).
\bibitem{C95}C. Ciofi degli Atti and S. Simula, Phys. Rev. {\bf C53} 1689 
(1996).
\bibitem{fantoni} S. Fantoni, private communication.
\bibitem{Mahaux}C. Mahaux and H. Ngo, Nucl. Phys. {\bf A378} (1981) 285.
\bibitem{cda_dat} C. Ciofi degli Atti, E. Pace and G. Salme, Phys. Rev. 
{\bf C43} (1991) 1153.


\end{thebibliography}
\end{document}